\documentclass[prb,twocolumn,aps,superscriptaddress,floatfix]{revtex4}
\usepackage{amsmath}
\usepackage{amssymb}
\usepackage{graphicx}
\usepackage{color}
\usepackage[svgnames]{xcolor}
\usepackage[normalem]{ulem}
\usepackage{cancel}
\usepackage{hyperref}
\usepackage{verbatim}

\begin{document}

\title{Ordering in SU(4)-symmetric model of AA~bilayer graphene}

\author{A.V. Rozhkov}
\affiliation{Institute for Theoretical and Applied Electrodynamics, Russian
Academy of Sciences, 125412 Moscow, Russia}

\author{A.O. Sboychakov}
\affiliation{Institute for Theoretical and Applied Electrodynamics, Russian
Academy of Sciences, 125412 Moscow, Russia}

\author{A.L. Rakhmanov}
\affiliation{Institute for Theoretical and Applied Electrodynamics, Russian
Academy of Sciences, 125412 Moscow, Russia}

\begin{abstract}
We examine possible ordered states of AA stacked bilayer graphene arising
due to electron-electron coupling. We show that under certain assumptions
the Hamiltonian of the system possesses an SU(4) symmetry. The
multicomponent order parameter is described by a
$4\times4$
matrix
$\hat{Q}$,
for which a mean-field self-consistency equation is derived. This equation
allows Hermitian and non-Hermitian solutions. Hermitian solutions can be
grouped into three topologically-distinct classes. First class corresponds
to the charge density wave. Second class includes spin density wave, valley
density wave, and spin-valley density wave. An ordered state in the third
class is a combination of all the aforementioned density-wave types. For
anti-Hermitian
$\hat{Q}$
the ordered states are characterized by spontaneous inter-layer loop
currents flowing in the bilayer. Depending on the topological class of the
solution these currents can carry charge, spin, valley, and spin-valley
quanta. We also discuss the special case when matrix
$\hat{Q}$
is not Hermitian and not anti-Hermitian. Utility and weak points of the
proposed SU(4)-based classification scheme of the ordered states are
analyzed.
\end{abstract}


\date{\today}

\maketitle

\section{Introduction}
\label{intro}

Discovery of series of alternating Mott insulating, metallic, and
superconducting states in the magic-angle twisted graphene
bilayer~\cite{twist_exp_insul2018,twist_exp_sc2018,lu2019ipsita}
was the first experimental evidence of the diversity of possible ordered
states in bilayer graphene systems caused by electron-electron coupling.
Recently, a cascade of transitions between different non-superconducting
and superconducting states was observed in the
well-researched~\cite{de2022cascade,zhou2022isospin,seiler2022quantum}
AB (or Bernal) stacked bilayer graphene (AB-BLG). This feature is not
limited to AB-BLG: the current studies
(theoretical~\cite{kokanova2021prb, sboychakov_FraM2021prb_lett},
numerical~\cite{corboz_rice2014tJcompetition,white_hubb_stripes2017numerics},
and experimental~\cite{narayanan2014coexistence})
indicate that even quite simple electronic systems may have several ordered
states competing against each other to become the true ground state. The
analysis of the cited above works shows that different ordered states of a
specific model are close to each other in terms of their (free) energy. As
a result, the true ground state of a system depends crucially on the
experimental conditions (temperature, magnetic field, pressure, sample
doping, substrates, etc.). Even a small change in any of these factors can
induce switching of the ground state. In such a situation, when a wide
class of the materials exhibit multiple transitions under weak variation of
the parameters, a convenient classification scheme of the possible ground
states is of significant help.

This paper is dedicated to theoretical study of an electronic liquid of AA
bilayer graphene (AA-BLG), a topic that attracted attention in recent
years~\cite{PrlOur, BreyFertig, PrbROur,
Sboychakov_PRB2013_PS_AAgraph, AkzyanovAABLG2014, Honerkamp,
aa_encapsulated2018theor, aa_electron_optics2020ab_init,
aa_order2021theor_apinyan}.
A specific question we would like to address here is the problem of
classifying low-temperature non-superconducting ordered states of AA-BLG.
The investigation of this relatively simple system can help us to extend
the proposed approach to other types of orders (superconducting, in
particular) and other types of systems, such as twisted and AB-BLG.

The present discussion is built on the approach previously used in
Ref.~\onlinecite{Nandkishore2010b}
to study multiple possible order parameters in the AB-BLG. Adopting that
technique for the AA-BLG electronic liquid, we will formulate an
approximate Hamiltonian that possesses an SU(4) symmetry in the spin-valley
index space at zero doping. Following
Ref.~\onlinecite{Nandkishore2010b},
we assume that the main interaction in the system is a long-range Coulomb
electron-electron coupling and neglect any additional interactions (e.g.,
electron-lattice). Mean field (matrix) self-consistency equation for such
an AA-BLG model reveals several competing non-superconducting ordered
states characterized by a matrix order parameter.

Studying the self-consistency equation, one finds that its solutions could
be both Hermitian and non-Hermitian matrices. The former case was
considered in Ref.~\onlinecite{Nandkishore2010b}. It has been shown that the signature of the order parameter matrix can be used for exhaustive classification of the ordered states.
Here we analyze also the non-Hermitian solutions to
the self-consistency equation. These solutions are characterized by finite
inter-layer currents that can carry, not
only electric charge, but also spin-related and/or valley-related quanta,
depending on the specific details.
A broad array of the ordered states compatible with our self-consistency
equation suggests that the true ground state of the AA-BLG may depend on
variety of details some of which could be
purposefully tailored to stabilize a desired order parameter.

The paper is organized as follows. In Sec.~II we formulate an
SU(4)-symmetric model of the AA-BLG.
In Sec.~III we derive the self-consistency mean
field equation for the multicomponent order parameter described by
$4\times4$ matrix $\hat{Q}$.
In Sec.~IV we describe the solutions of this mean field equations. Three
different cases are considered: Hermitian
$\hat{Q}$,
anti-Hermitian
$\hat{Q}$,
and matrix $\hat{Q}$ that is neither Hermitian nor anti-Hermitian (non-Hermitian
and non-anti-Hermitian $\hat{Q}$).
Section~V is devoted to discussion. Some details of derivation of the
mean field equation are placed in the Appendix.

\section{Model}
\label{model}

\subsection{Tight-binding kinetic energy for AA-BLG}
\label{subsect::TB}

A sample of the AA-BLG consists of two graphene layers and every carbon atom of the top layer is directly above one of the carbon atoms in the bottom layer. Single-electron hopping Hamiltonian for the AA-BLG can be written as~\cite{Sboychakov_PRB2013_PS_AAgraph}
\begin{eqnarray}
\label{SingleParticleHam}
\hat{H}_0
=
-t\sum_{\langle \mathbf{mn}\rangle l\sigma}
        \left(
		\hat{d}^\dag_{\mathbf{m}lA\sigma}\hat{d}_{\mathbf{n}lB\sigma}
		+
		{\rm h.c.}
	\right)
\\
\nonumber
-t_0\sum_{\mathbf{n}a\sigma}
        \left(
		\hat{d}^\dag_{\mathbf{n}1a\sigma}\hat{d}_{\mathbf{n}2a\sigma}
		+
		{\rm h.c.}
	\right).
\end{eqnarray}
Here
$\hat{d}\,^\dag_{\mathbf{n}la\sigma}$
and
$\hat{d}_{\mathbf{n}la\sigma}$
are the creation and annihilation operators of an electron with spin
projection $\sigma$ in the layer
$l = 1,2$
on the sublattice
$a = A,B$
at the unit cell
$\mathbf{n}$;
and
$\langle...\rangle$
denotes a nearest-neighbor pair. The amplitude
$t = 2.7$\,eV
($t_0 = 0.35$\,eV)
in
Eq.~\eqref{SingleParticleHam}
describes the in-plane (inter-plane) nearest-neighbor hopping.

\begin{figure}
\centering
\includegraphics[width=0.95\columnwidth]{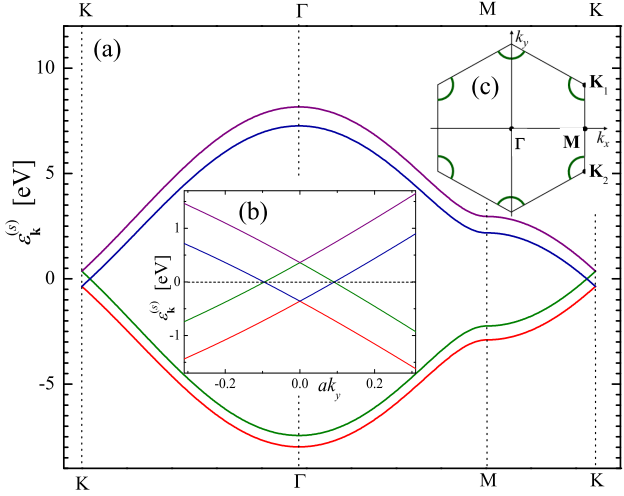}\\
\caption{(Color online) (a) The single-particle band structure of the
AA-BLG. The spectrum consists of four bands
$\epsilon^{(s)}_{\bf k}$,
see
Eqs.~(\ref{eq::TB_spectrum_12})
and~(\ref{eq::TB_spectrum_34}).
(b) The spectrum near the Dirac points can be approximately linearized, see
Eqs.~(\ref{eq::TB_spectrum_12_approx})
and~(\ref{eq::TB_spectrum_34_approx}).
The intersection of the bands $s=2$ and $s=3$
occurs exactly at zero energy, which corresponds to the Fermi level of the
undoped system. (c) The first Brillouin zone (hexagon) of the AA-BLG. The circles around
the Dirac points correspond to Fermi surfaces.
\label{fig::BZ_graphene}
}
\end{figure}

To diagonalize the hopping Hamiltonian it is convenient to switch to
momentum representation. For a sample with
$N_c$
unit cells in a single graphene layer this is achieved by the Fourier
transformation
\begin{eqnarray}
\label{eq::def_Fourier}
\hat{d}_{{\bf k} la \sigma}
=
\frac{e^{- ia\varphi_{\bf k} }}{\sqrt{N_c}}
\sum_{\bf n}
	e^{i {\bf k \cdot n}} \hat{d}_{{\bf n}la \sigma}.
\end{eqnarray}
Here the numerical values of the sublattice index $a$ are
$a=0$
for sublattice $A$, and
$a=1$
for sublattice $B$. The phase factor in
Eq.~(\ref{eq::def_Fourier})
is
$\exp(i \varphi_{\bf k})
= f_{\bf k}/|f_{\bf k}|$,
the function
$f_{\bf k}$
being
\begin{equation}
\label{f_k}
f_\mathbf{k}
=
1+2\exp{\left(\frac{3ik_xa_0}{2}\right)}
\cos{\left(\frac{\sqrt{3}k_ya_0}{2}\right)},
\end{equation}
and
$a_0$
is the in-plane carbon-carbon distance. Quasi-momentum vectors are confined
to the graphene Brillouin zone, which has a shape of a regular hexagon with two independent Dirac
points in two corners~\cite{ourBLGreview2016}
\begin{eqnarray}
\mathbf{K}_{1,2} = \frac{2\pi}{3\sqrt{3}a_0}(\sqrt{3}, \pm 1).
\end{eqnarray}
Four more corners of the Brillouin zone can be found by two
$60^\circ$
rotations of
$\mathbf{K}_{1,2}$,
see
Fig.~\ref{fig::BZ_graphene}.

In the momentum representation, the hopping Hamiltonian becomes
\begin{eqnarray}
\hat{H}_0
=
\sum_{{\bf k} \sigma}
	\hat{\Psi}_{{\bf k} \sigma}^\dag
	\hat{\cal H}_{\bf k}^{\vphantom{\dagger}}
	\hat{\Psi}_{{\bf k} \sigma}^{\vphantom{\dagger}},
\end{eqnarray}
where the matrix
$\hat{\cal H}_{\bf k}$
and the bi-spinor
$\hat{\Psi}_{{\bf k} \sigma}^\dag$
are
\begin{eqnarray}
\hat{\cal H}_{\bf k}^{\vphantom{\dagger}}
=
- \left( \begin{matrix}
	0 & t_0 & t|f_{\bf k}| & 0\cr
	t_0 & 0 & 0 & t|f_{\bf k}| \cr
	t|f_{\bf k}|& 0 & 0 & t_0 \cr
	0 & t|f_{\bf k}| & t_0 & 0 \cr
\end{matrix}\right),
\\
\hat{\Psi}_{{\bf k} \sigma}^\dag
=
\left(
	\hat{d}_{{\bf k} 1 A \sigma}^\dag,
	\hat{d}_{{\bf k} 2 A \sigma}^\dag,
	\hat{d}_{{\bf k} 1 B \sigma}^\dag,
	\hat{d}_{{\bf k} 2 B \sigma}^\dag
\right).
\end{eqnarray}
Thus the
Hamiltonian~(\ref{SingleParticleHam})
can be diagonalized as
\begin{eqnarray}
\label{eq::HamDiag}
\hat{H}_0
=
\sum_{\mathbf{k}s\sigma}
        \epsilon^{(s)}_{\mathbf{k}}
	\hat{\gamma}^\dag_{\mathbf{k}s\sigma}
	\hat{\gamma}_{\mathbf{k}s\sigma}^{\vphantom{\dagger}},
\end{eqnarray}
where the band eigenenergies
$\epsilon^{(s)}_{\mathbf{k}}$
are
\begin{eqnarray}
\label{eq::TB_spectrum_12}
\epsilon^{(1)}_{\mathbf{k}}
&=&
-t_0-t|f_{\bf k}|,
\quad
\epsilon^{(2)}_{\mathbf{k}}= -t_0+t|f_{\bf k}|,
\\
\label{eq::TB_spectrum_34}
  \epsilon^{(3)}_{\mathbf{k}} &=& +t_0-t|f_{\bf k}|,
\quad
\epsilon^{(4)}_{\mathbf{k}}=+t_0+t|f_{\bf k}|.
\end{eqnarray}
This energy spectrum is plotted in Fig.~\ref{fig::BZ_graphene}(a).

The band operators
$\hat{\gamma}_{{\bf k} s \sigma}$
are connected to
$\hat{d}_{{\bf k} la \sigma}$
as follows
\begin{eqnarray}
\label{eq::gamma_d1}
\hat{d}_{\mathbf{k}la\sigma}
=
\frac{1}{2} \left[
	\hat{\gamma}_{\mathbf{k}1\sigma}
	+
	(-1)^a\hat{\gamma}_{\mathbf{k}2\sigma}
	+
	(-1)^l\hat{\gamma}_{\mathbf{k}3\sigma}
\right.
\\
\nonumber
\left.
	+
	(-1)^{a+l}\hat{\gamma}_{\mathbf{k}4\sigma}
\right],
\end{eqnarray}
where
$l=0$
for layer 1 and
$l=1$
for layer 2. The latter relation is easy to invert and find that
\begin{eqnarray}
\label{eq::gamma12_def}
\hat{\gamma}_{\mathbf{k}1\sigma}
=
\frac{1}{2} \sum_{la} \hat{d}_{\mathbf{k}la\sigma},
\quad
\hat{\gamma}_{\mathbf{k}2\sigma}
=
\frac{1}{2} \sum_{la} (-1)^a \hat{d}_{\mathbf{k}la\sigma},
\\
\label{eq::gamma34_def}
\hat{\gamma}_{\mathbf{k}3\sigma}
=
\frac{1}{2} \sum_{la} (-1)^l \hat{d}_{\mathbf{k}la\sigma},
\
\hat{\gamma}_{\mathbf{k}4\sigma}
=
\frac{1}{2} \sum_{la} (-1)^{l+a} \hat{d}_{\mathbf{k}la\sigma}.
\end{eqnarray}
Analyzing
spectra~(\ref{eq::TB_spectrum_12})
and~(\ref{eq::TB_spectrum_34})
one notices that the bands
$s = 2, 3$
cross the Fermi level
$\varepsilon = 0$.
The corresponding Fermi surfaces can be approximated by circles of radius
$k_{\rm F} = 2t_0/(3 t a_0)$
centered around the Brillouin zone corners (the Dirac points), as shown in
Fig.~\ref{fig::BZ_graphene}\,(c).
At the same time, the bands
$s = 1, 4$
do not reach the Fermi level, and have no Fermi surface.

\subsection{Valley quantum number}

For graphene-based systems it is often useful to introduce a binary-valued
valley index $\xi = 1,2$: an electronic state with the quasi-momentum
${\bf k}$ is assumed to be in valley
${\bf K}_\xi$
if
$| {\bf k} - {\bf K}_\xi | < q_0$,
where the valley radius
$q_0=|{\bf K}_1-{\bf K}_2|/2$
equals
$q_0 = 2\pi/3\sqrt{3}a_0$.
The states whose momenta lie outside either valley
${\bf K}_1$
or valley
${\bf K}_2$
are high-energy states. Such states will be discarded since their
contribution to the low-energy physics is insignificant. Further we will
count the quasi-momentum
${\bf k}$
relative to the valley centers 
${\bf K}_{1,2}$.
We expand the function
$f_{{\bf K}_\xi + {\bf k}}$
near each Dirac point and in the linear approximation obtain
\begin{equation}
\label{eq::linear_f_k}
f_{{\bf K}_\xi + {\bf k}}
=
\frac{3a_0}{2}\left[k_y+(-1)^\xi ik_x\right].
\end{equation}
Since we are interested here only in the low-energy states, we will use linear approximation~\eqref{eq::linear_f_k} within the valleys.

We define new single-electron operators in a specific valley as
\begin{eqnarray}
\label{eq::gamma_valley}
\hat{\gamma}_{\mathbf{k}s \xi \sigma}
=
\begin{cases}
	\hat{\gamma}_{{\bf K}_1 + \mathbf{k}s \sigma}, &
				\text{if } \xi = 1, \\
	(-1)^{s+1} e^{i \phi_{\bf k} }
	\hat{\gamma}_{{\bf K}_2 + \mathbf{k}s \sigma}, &
				\text{if } \xi = 2,
\end{cases}
\end{eqnarray}
where the phase factor $\exp(i \phi_{\bf k})$ is equal to
\begin{eqnarray}
e^{i \phi_{\bf k}} =
- \frac{ i k_x + k_y }{|{\bf k}|}.
\end{eqnarray}
As it follows from Eq.~\eqref{eq::linear_f_k}, it is connected with the complex phases near the valley centers
${\bf K}_{1,2}$
\begin{eqnarray}
\label{eq::phase_relation}
e^{i \varphi_{{\bf K}_1 + {\bf k} }} = - e^{- i \phi_{\bf k}},
\quad
e^{i \varphi_{{\bf K}_2 + {\bf k} }} = e^{i \phi_{\bf k}}.
\end{eqnarray}
Using
Eqs.~(\ref{eq::gamma_d1})
and~(\ref{eq::gamma_valley}),
one can write
$\hat{d}_{\mathbf{n} la \sigma}$
in terms of
$\hat{\gamma}_{\mathbf{k}s \xi \sigma}$.
To this purpose, it is convenient to introduce the valley-specific operator
$\hat{d}_{\mathbf{k}la \xi \sigma}
=
\hat{d}_{{\bf K}_\xi+\mathbf{k}la \sigma}$,
and write
\begin{eqnarray}
\label{eq::d_valley}
\hat{d}_{\mathbf{k}la \xi \sigma}
=
\begin{cases}
	\hat{\mathfrak{g}}_{{\bf k} la \xi \sigma}, &
		\text{if } \xi=1, \\
	e^{-i\phi_{\bf k}} \hat{\mathfrak{g}}_{{\bf k}l\bar{a}\xi\sigma}, &
		\text{if } \xi=2.
\end{cases}
\end{eqnarray}
In this definition, the operators
$\hat{\mathfrak{g}}_{{\bf k}l a \xi\sigma}$
are linear combinations of the band operators
\begin{eqnarray}
\label{eq::GAMMA_valley}
\hat{\mathfrak{g}}_{{\bf k}l a \xi\sigma}
=
\frac{1}{2} \left[
	\hat{\gamma}_{{\bf k} 1 \xi\sigma}
	+
	(-1)^a \hat{\gamma}_{{\bf k} 2 \xi\sigma}
	+
	(-1)^l \hat{\gamma}_{{\bf k} 3 \xi\sigma}
\right.
\\
\nonumber
\left.
	+
	(-1)^{a+l} \hat{\gamma}_{{\bf k} 4 \xi\sigma}
\right],
\end{eqnarray}
and we adhere to the convention that a bar over a binary-valued index
inverts its value (i.e., if
$a = A$,
then
$\bar{a} = B$,
and vice versa).

There is an obvious disparity between the valleys in
definitions~(\ref{eq::gamma_valley})
and~(\ref{eq::d_valley}).
Note that the sublattice index in
Eq.~(\ref{eq::d_valley})
is inverted for
$\xi = 2$.
In addition, the phase factors
in~(\ref{eq::gamma_valley})
are not identical in different valleys. The same is true for
Eq.~(\ref{eq::d_valley}).
We will see below that such a phase factor choice is needed to make explicit the SU(4) symmetry of the interaction term.

Inverting 
relation~(\ref{eq::def_Fourier})
and using the valley-specific operators in
${\bf k}$-space,
we can approximate the real-space operators as follows
\begin{eqnarray}\label{real_space_op}
\hat{d}_{{\bf n} la \sigma}
\approx
\frac{1}{\sqrt{N_c}}
\sum_{|{\bf k}|<q_0}
	\left[
		e^{-ia \phi_{\bf k} - i {\bf K}_1 \cdot {\bf n}}
		(-1)^a\hat{\mathfrak{g}}_{{\bf k} la {\bf K}_1 \sigma}
	\right.
\\
\nonumber
	\left.
		+
		e^{-i \bar{a} \phi_{\bf k} - i {\bf K}_2 \cdot {\bf n}}
		\hat{\mathfrak{g}}_{{\bf k} l\bar{a} {\bf K}_2 \sigma}
	\right]
	e^{- i {\bf k} \cdot {\bf n}}.
\end{eqnarray}
This expression disregards all high-energy states that lies outside the
valleys.

\subsection{SU(4)-symmetric single-electron Hamiltonian}

Within the developed formalism the Hamiltonian~(\ref{eq::HamDiag})
can be approximated as
\begin{eqnarray}
\label{eq::HamDiag_valley}
\hat{H}_0
\approx
\sum_{ s , |{\bf k}|<q_0}
        \varepsilon^{(s)}_{\mathbf{k}}
\sum_{ \xi \sigma}
        \hat{\gamma}^\dag_{\mathbf{k}s \xi\sigma}
        \hat{\gamma}_{\mathbf{k}s \xi \sigma}^{\vphantom{\dagger}},
\end{eqnarray}
where
$\varepsilon^{(s)}_{\mathbf{k}}$
are linear approximations to the exact eigenenergies
$\epsilon^{(s)}_{\mathbf{k}}$
near the Dirac points
\begin{eqnarray}
\label{eq::TB_spectrum_12_approx}
\varepsilon^{(1)}_{\mathbf{k}} = -t_0-v_{\rm F} |{\bf k}|,
\quad
\varepsilon^{(2)}_{\mathbf{k}}= -t_0+v_{\rm F} |{\bf k}|,
\\
\label{eq::TB_spectrum_34_approx}
\varepsilon^{(3)}_{\mathbf{k}} = +t_0-v_{\rm F}|{\bf k}|,
\quad
\varepsilon^{(4)}_{\mathbf{k}}=+t_0+v_{\rm F}|{\bf k}|.
\end{eqnarray}
The Fermi velocity in these expressions is equal to
$v_{\rm F} = 3 a_0 t/2$ ($\hbar=1$).

The significance of
formula~(\ref{eq::HamDiag_valley})
is that it explicitly demonstrates the valley degeneracy of the
single-electron spectrum of AA-BLG, and, additionally,
it reveals the SU(4) symmetry of the model. To illustrate this important
point we introduce the spin-valley muti-index
$m=(\xi, \sigma)$,
which takes four possible values. This allows us to abbreviate the notation
as follows
$\hat{\gamma}_{\mathbf{k}s \xi \sigma} = \hat{\gamma}_{\mathbf{k}s m}$.
Any 4$\times$4 unitary matrix
$\hat{\cal U} \in$~SU(4),
with matrix elements
$u_{mm'}$,
defines a Bogolyubov transform
\begin{eqnarray}
\label{eq::Bogolyubov_gamma}
\hat{\gamma}_{{\bf k} s m}
\rightarrow
\sum_{m'} u_{mm'} \hat{\gamma}_{\mathbf{k}s m'}.
\end{eqnarray}
It is easy to check that this transformation leaves
Hamiltonian~(\ref{eq::HamDiag_valley})
unchanged.

\subsection{Interaction term}

The most general form of the interaction term for AA-BLG is
\begin{eqnarray}
\label{interaction_q}
\hat{H}_{\textrm{int}}
=
\frac{1}{2 N_c}
\sum_{\mathbf{k}ll'aa'}
        V_{{\bf k} aa'}^{ll'} \hat{\rho}_{{\bf k} l a} \hat{\rho}_{-{\bf k} l' a'}.
\end{eqnarray}
Vector
${\bf k}$
here is the transferred momentum, parameters
$V_{{\bf k} aa'}^{ll'}$
are the Fourier components of the potential energy
$V_{aa'}^{ll'} ({\bf R})$
describing the interaction between an electron in layer $l$, sublattice $a$
and another electron in layer $l'$, sublattice $a'$. Finally,
$\hat{\rho}_{{\bf k} l a}$
is the Fourier component of a single-site particle-density operator
$\hat{\rho}_{ {\bf n} l a} = \sum_\sigma
	\hat{d}^\dag_{\mathbf{n}la\sigma}
	\hat{d}_{\mathbf{n}la\sigma}^{\vphantom{\dagger}}$.

For small transferred momentum
${\bf k}$
one has
$\hat{\rho}_{{\bf k} l a}
=
\hat{\rho}_{{\bf k} l a}^{{\bf K}_1} + \hat{\rho}_{{\bf k} l a}^{ {\bf K}_2}$,
where two chiral density components can be expressed as
\begin{eqnarray}
\hat{\rho}_{{\bf k} l a}^\xi
=
\sum_{{\bf q} \sigma}
	e^{i a (\varphi_{{\bf K}_\xi + {\bf k+q}}
		-
		\varphi_{{\bf K}_\xi + {\bf q}} )}
	\hat{d}^\dag_{{\bf q} l a \xi \sigma}
	\hat{d}^{\vphantom{\dag}}_{{\bf k} + {\bf q} l a \xi \sigma},
\end{eqnarray}
or, equivalently, in terms of the band operators as
\begin{eqnarray}
\hat{\rho}_{{\bf k} l a}^{{\bf K}_1}
=
\sum_{{\bf q} \sigma}
	e^{- i a (\phi_{\bf k+q} - \phi_{\bf q} )}
	\hat{\mathfrak{g}}^\dag_{{\bf q} l a {\bf K}_1 \sigma}
	\hat{\mathfrak{g}}^{\vphantom{\dag}}_{{\bf k} + {\bf q} l a {\bf K}_1 \sigma},
\\
\hat{\rho}_{{\bf k} l a}^{{\bf K}_2}
=
\sum_{{\bf q} \sigma}
	e^{- i \bar{a} (\phi_{\bf k+q} - \phi_{\bf q} )}
	\hat{\mathfrak{g}}^\dag_{{\bf q} l \bar{a} {\bf K}_2 \sigma}
	\hat{\mathfrak{g}}^{\vphantom{\dag}}_{{\bf k} + {\bf q} l \bar{a} {\bf K}_2 \sigma}.
\end{eqnarray}
Both
$\hat{\rho}_{{\bf k} l a}^{\xi}$
vary smoothly in space for small
${\bf k}$.
Besides
$\hat{\rho}_{{\bf k} l a}^{\xi}$,
there is an oscillating contribution to the density
\begin{eqnarray}
\hat{\rho}_{{\bf k} l a}^{\xi\bar{\xi}}
=
\sum_{{\bf q} \sigma}
	e^{i a (\varphi_{\bf K_\xi+ k+q} - \varphi_{\bf K_{\bar{\xi}}+q} )}
	\hat{d}^\dag_{{\bf q} l a \bar{\xi} \sigma}
	\hat{d}^{\vphantom{\dag}}_{{\bf k} + {\bf q} l a \xi \sigma}.
\end{eqnarray}
The wave vector corresponding to the spatial modulation of
$\hat{\rho}_{{\bf k} l a}^{\xi\bar{\xi}}$
is never small: it is of order of
${\bf K}_1 - {\bf K}_2$
even for small
${\bf k}$.

Since the density operator has smooth as well as oscillating contributions,
the interaction can be split into the forward-scattering
$\hat{H}_{\textrm{f}}$ and back-scattering $\hat{H}_{\textrm{b}}$
\begin{eqnarray}
\hat{H}_{\textrm{int}}
=
\hat{H}_{\textrm{f}}
+
\hat{H}_{\textrm{b}},
\\
\hat{H}_{\textrm{f}}
=
\frac{1}{2 N_c}
\sum_{\mathbf{k}\xi \xi' \atop ll' aa'}
        V_{{\bf k} aa'}^{ll'} \hat{\rho}^\xi_{{\bf k} l a} \hat{\rho}^{\xi'}_{-{\bf k} l' a'}.
\\
\hat{H}_{\textrm{b}}
=
\frac{1}{2 N_c}
\sum_{\mathbf{k}\xi ll'aa'}
        V_{{\bf K}_1 - {\bf K}_2 aa'}^{ll'}
	\hat{\rho}^{\xi\bar{\xi}}_{{\bf k} l a}
	\hat{\rho}^{\bar{\xi}\xi}_{-{\bf k} l' a'}.
\end{eqnarray}
As one can see,
$\hat{H}_{\textrm{f}}$
describes scattering in which both participating electrons maintain their
valley indices after scattering, while
$\hat{H}_{\textrm{b}}$
represents large-momentum scattering, when participating electrons from
different valleys exchange their valley indices.

We assume here that the electron-electron interaction is sufficiently long-range. In this case $V_{{\bf K}_1 - {\bf K}_2 aa'}^{ll'} < V_{{\bf k} aa'}^{ll'}$ since
$|{\bf k}| < |{\bf K}_1 - {\bf K}_2|$. For this reason we neglect below the
back-scattering (this issue will be discussed in
Sec.~\ref{sec::Discussion}
in more detail). Since the considered interaction is a long-range one, we
can assume that the coupling is approximately independent of the sublattice
indices:
\begin{eqnarray}
V_{{\bf k} aa'}^{ll'} \approx V_{\bf k}^{ll'}.
\end{eqnarray}
Under these approximations, the interaction Eq.~\eqref{interaction_q} reads
\begin{eqnarray}
\label{eq::interaction_f}
\hat{H}_{\textrm{int}}
=
\frac{1}{2 N_c}
\sum_{\mathbf{k} ll'}
        V_{{\bf k} }^{ll'} \hat{\rho}_{{\bf k} l} \hat{\rho}_{-{\bf k} l'},
\end{eqnarray}
where the smooth density component in layer $l$ is
\begin{eqnarray}
\label{eq::rho_invariant}
\hat{\rho}_{{\bf k} l } = \sum_{a \xi } \hat{\rho}_{{\bf k} l a}^\xi
=
\sum_{{\bf q} a m}
	e^{i a (\phi_{\bf q} - \phi_{\bf k+q} )}
	\hat{\mathfrak{g}}^\dag_{{\bf q} l a m}
	\hat{\mathfrak{g}}^{\vphantom{\dag}}_{{\bf k} + {\bf q} l a m}.
\end{eqnarray}
In the latter formula we used the multi-index notation
$m=(\xi, \sigma)$.
This serves twofold purpose. For one, it makes the expression more concise.
Additionally, it explicitly reveals the invariance of
$\hat{\rho}_{{\bf k} l }$
under the action of the SU(4) Bogolyubov
transformation~(\ref{eq::Bogolyubov_gamma}).
At this point one can appreciate the motivation behind the complexity of
formulas~(\ref{eq::gamma_valley})
and~(\ref{eq::d_valley}).
If multiple phase factors were not absorbed in the definitions of the
valley-specific operators, these phase factors would emerge in
Eq.~(\ref{eq::rho_invariant}),
obscuring the invariance. Finally, we observe that, since the operator
$\hat{\rho}_{{\bf k} l }$
possesses the SU(4) invariance, the same is true for the
interaction~(\ref{eq::interaction_f}).

\subsection{Effective model}

As it was stated above, only two of the four single-electron bands form the
Fermi surface at zero doping (see
Fig.~\ref{fig::BZ_graphene}).
Therefore, the high-energy bands
$s=1,4$
can not modify significantly the low-energy physics of the AA-BLG. We
discarded these bands from the model, which simplifies considerably further
analysis. In this approximation the single-electron Hamiltonian becomes
\begin{eqnarray}
\label{eq::effective_kinetic}
\hat{H}_0^{\rm eff}
=
\sum_{\mathbf{k}\xi\sigma}
        (v_{\rm F} |{\bf k}| - t_0)
	(\hat{\gamma}^\dag_{\mathbf{k}2\xi\sigma}
	\hat{\gamma}_{\mathbf{k}2\xi\sigma}^{\vphantom{\dagger}}
	-
	\hat{\gamma}^\dag_{\mathbf{k}3\xi\sigma}
	\hat{\gamma}_{\mathbf{k}3\xi\sigma}^{\vphantom{\dagger}}).
\end{eqnarray}
The density operator $\hat{\rho}_{{\bf k} l}$ reduces to
\begin{eqnarray}
\hat{\rho}_{{\bf k} l }
\approx
\frac{1}{4}
\sum_{{\bf q} a m}
	e^{ i a (\phi_{\bf q} - \phi_{\bf q+k} )}
	[\hat{\gamma}^\dag_{{\bf q} 2 m}
	+
	(-1)^{l+a} \hat{\gamma}^\dag_{{\bf q} 3 m}]
\\
\nonumber
\times
	[\hat{\gamma}^{\vphantom{\dag}}_{{\bf q} + {\bf k} 2 m}
	+
	(-1)^{l+a} \hat{\gamma}^{\vphantom{\dag}}_{{\bf q} + {\bf k} 3 m}].
\end{eqnarray}
Substituting this expression in Eq.~(\ref{eq::interaction_f}) one derives
\begin{eqnarray}
\hat{H}^{\rm eff}_{\rm int}
=
\hat{H}^{\rm eff}_{\rm dir}
+
\hat{H}^{\rm eff}_{\rm ex}
+
\hat{H}^{\rm eff}_{\rm u},
\end{eqnarray}
where the direct term is defined as
\begin{eqnarray}
\hat{H}^{\rm eff}_{\rm dir}
=
\frac{1}{16 N_c}
\sum_{{\bf q} {\bf q}' {\bf k} m m'}
	V_{+} ({\bf k})
	\left[1+ e^{ i (\phi_{\bf q} - \phi_{\bf q+k} )} \right]
\\
\nonumber
\times
	\left[ 1+ e^{ i (\phi_{\bf q'} - \phi_{\bf q'-k} )} \right]
	\left(
		\hat{\gamma}^\dag_{{\bf q} 2 m}
		\hat{\gamma}^{\vphantom{\dag}}_{{\bf q} + {\bf k} 2 m}
		+
		\hat{\gamma}^\dag_{{\bf q} 3 m}
		\hat{\gamma}^{\vphantom{\dag}}_{{\bf q} + {\bf k} 3 m}
	\right)
\\
\nonumber
\times
	\left(
		\hat{\gamma}^\dag_{{\bf q}' 2 m'}
		\hat{\gamma}^{\vphantom{\dag}}_{{\bf q}' - {\bf k} 2 m'}
		+
		\hat{\gamma}^\dag_{{\bf q}' 3 m'}
		\hat{\gamma}^{\vphantom{\dag}}_{{\bf q}' - {\bf k} 3 m'}
	\right),
\end{eqnarray}
the exchange term is
\begin{eqnarray}
\label{eq::exchange}
\hat{H}^{\rm eff}_{\rm ex}
=
\frac{1}{16 N_c}
\sum_{{\bf q} {\bf q}' {\bf k} m m'}
	V_{-} ({\bf k})
	\left[1- e^{ i (\phi_{\bf q} - \phi_{\bf q+k} )} \right]
\\
\nonumber
\times
	\left[ 1- e^{ i (\phi_{\bf q'} - \phi_{\bf q'-k} )} \right]
	\left(
		\hat{\gamma}^\dag_{{\bf q} 2 m}
		\hat{\gamma}^{\vphantom{\dag}}_{{\bf q} + {\bf k} 3 m}
		\hat{\gamma}^\dag_{{\bf q}' 3 m'}
		\hat{\gamma}^{\vphantom{\dag}}_{{\bf q}' - {\bf k} 2 m'}
	\right.
\\
\nonumber
	\left.
		+
		\hat{\gamma}^\dag_{{\bf q} 3 m}
		\hat{\gamma}^{\vphantom{\dag}}_{{\bf q} + {\bf k} 2 m}
		\hat{\gamma}^\dag_{{\bf q}' 2 m'}
		\hat{\gamma}^{\vphantom{\dag}}_{{\bf q}' - {\bf k} 3 m'}
	\right),
\end{eqnarray}
and the umklapp term is
\begin{eqnarray}
\hat{H}^{\rm eff}_{\rm u}
=
\frac{1}{16 N_c}
\sum_{{\bf q} {\bf q}' {\bf k} m m'}
	V_{-} ({\bf k})
	\left[1- e^{ i (\phi_{\bf q} - \phi_{\bf q+k} )} \right]
\\
\nonumber
\times
	\left[ 1- e^{ i (\phi_{\bf q'} - \phi_{\bf q'-k} )} \right]
	\left(
		\hat{\gamma}^\dag_{{\bf q} 2 m}
		\hat{\gamma}^{\vphantom{\dag}}_{{\bf q} + {\bf k} 3 m}
		\hat{\gamma}^\dag_{{\bf q}' 2 m'}
		\hat{\gamma}^{\vphantom{\dag}}_{{\bf q}' - {\bf k} 3 m'}
	\right.
\\
\nonumber
	\left.
		+
		\hat{\gamma}^\dag_{{\bf q} 3 m}
		\hat{\gamma}^{\vphantom{\dag}}_{{\bf q} + {\bf k} 2 m}
		\hat{\gamma}^\dag_{{\bf q}' 3 m'}
		\hat{\gamma}^{\vphantom{\dag}}_{{\bf q}' - {\bf k} 2 m'}
	\right).
\end{eqnarray}
In these approximate expressions we introduced layer-symmetric and
layer-antisymmetric interactions
$V_{\pm} ({\bf k})
= \left(V_{\bf k}^{11} \pm V_{\bf k}^{12} \right)$.
We also used the relation
$\sum_{ll'} (-1)^l V^{ll'}_{\bf k} \equiv 0$,
which can be trivially checked.

\subsection{Symmetry group of the effective model}

Observe that the operators
$\hat{H}^{\rm eff}_0$,
$\hat{H}^{\rm eff}_{\rm dir}$,
$\hat{H}^{\rm eff}_{\rm ex}$,
and
$\hat{H}^{\rm eff}_{\rm u}$,
which constitute the effective model Hamiltonian, are individually
SU(4)-invariant. Indeed, each of these operators are explicitly composed of
the bilinears
${\cal I}^{sr}_{{\bf q} {\bf p}}$
defined as
\begin{eqnarray}
\label{eq::scalar_I}
{\cal I}^{sr}_{{\bf q} {\bf p}}
=
\sum_m \hat{\gamma}^\dag_{{\bf q} s m}
\hat{\gamma}^{\vphantom{\dag}}_{{\bf p} r m},
\quad
s,r = 2,3,
\end{eqnarray}
that are invariants of the SU(4) Bogolyubov
transformation~(\ref{eq::Bogolyubov_gamma}). Besides, the effective Hamiltonian evidently remains unchanged upon the substitution
\begin{eqnarray}
\label{eq::Z2}
\hat{\gamma}_{{\bf p} s m} \rightarrow (-1)^s \hat{\gamma}_{{\bf p} s m}.
\end{eqnarray}
This allows us to change the relative sign between
$\hat{\gamma}_{{\bf p} 3 m}$
and
$\hat{\gamma}_{{\bf p} 2 m}$,
without changing the Hamiltonian.

From the standpoint of the AA-BLG lattice structure, the
substitution~(\ref{eq::Z2}) corresponds to either switching the layers
\begin{eqnarray}
{\rm (top \ layer)} \leftrightarrow {\rm (bottom \ layer)},
\end{eqnarray}
or switching the sublattices
\begin{eqnarray}
{\rm (sublattice}\ A) \leftrightarrow {\rm (sublattice}\ B),
\end{eqnarray}
as it follows from relation~(\ref{eq::gamma_d1}).
Consequently, the
symmetry~(\ref{eq::Z2})
can be viewed as a manifestation of the layer equivalence, or manifestation
of the sublattice equivalence (at the level of our effective model these
two equivalences cannot be distinguished).

Transformation~(\ref{eq::Z2}),
together with the identity transformation, constitutes the
$\mathbb{Z}_2$
group. Since the
transformations~(\ref{eq::Bogolyubov_gamma})
and~(\ref{eq::Z2})
commute with each other, we conclude that the symmetry group of the effective Hamiltonian is
\begin{eqnarray}
G \cong {\rm SU}(4)\times \mathbb{Z}_2.
\end{eqnarray}
However, the symmetry group of $\hat{H}^{\rm eff}_0$ and $\hat{H}^{\rm eff}_{\rm dir}$
is broader than $G$: one can check directly that both $\hat{H}^{\rm eff}_0$ and $\hat{H}^{\rm eff}_{\rm dir}$ are composed of
the bilinears
${\cal I}^{sr}_{{\bf q} {\bf p}}$
with identical band indices
$s=r$.
Thus,
$\hat{H}^{\rm eff}_0$
and
$\hat{H}^{\rm eff}_{\rm dir}$
remain invariant even when a Bogolyubov rotation for
$s=2$
band is non-identical to the rotation for
$s=3$
band.
In other words, when
$V_- \equiv 0$,
the model's symmetry group expands to
\begin{eqnarray}
G_0 \cong {\rm SU}(4) \times {\rm SU}(4).
\end{eqnarray}
We will see below that the broader symmetry group corresponds to broader
set of solutions for a self-consistency equation.

\section{Mean field approximation}

We apply the mean field approach to the effective Hamiltonian
\begin{eqnarray}
\hat{H}^{\rm eff}
=
\hat{H}_0^{\rm eff}
+
\hat{H}_{\rm int}^{\rm eff}.
\end{eqnarray}
with the aim of exploring (non-superconducting) symmetry breaking ordered phases of our model.
Implementing the
mean field decoupling for
$\hat{H}^{\rm eff}_{\rm dir}$,
we obtain
\begin{eqnarray}
\hat{H}^{\rm MF}_{\rm dir}
=
- \frac{1}{8 N_c}
\sum_{{\bf q} {\bf p} \atop m m'}
	V_+ ({\bf p} - {\bf q})
	\left|1+ e^{ i (\phi_{\bf q} - \phi_{{\bf p}} )} \right|^2
\\
\nonumber
\times\!\!
	\left(\!
		\langle \hat{\gamma}^\dag_{{\bf p} 3 m'}
			\hat{\gamma}^{\vphantom{\dag}}_{{\bf p}  2 m}
		\rangle
		\hat{\gamma}^\dag_{{\bf q} 2 m}
		\hat{\gamma}^{\vphantom{\dag}}_{{\bf q} 3 m'}
		+
		\hat{\gamma}^\dag_{{\bf p} 3 m'}
		\hat{\gamma}^{\vphantom{\dag}}_{{\bf p}  2 m}
		\langle
			\hat{\gamma}^\dag_{{\bf q} 2 m}
			\hat{\gamma}^{\vphantom{\dag}}_{{\bf q} 3 m'}
		\rangle\!
	\right)\!,
\end{eqnarray}
where
$\langle...\rangle$
stands for the ground-state average.

It is convenient to introduce the $4\times4$ operator-valued matrix
$\hat{\Theta}_{\bf q}$
whose elements are
\begin{eqnarray}
\Theta_{{\bf q} mm'}
=
\hat{\gamma}_{\mathbf{q} 3 m}^\dag
 \hat{\gamma}_{\mathbf{q}2 m'}^{\vphantom{\dagger}}.
\end{eqnarray}
Assuming that the average
$\langle\hat{\Theta}_{\bf q}\rangle$
depends only on the absolute value of the vector
$\mathbf{q}$,
we write the following compact expression
\begin{eqnarray}
\hat{H}^{\rm MF}_{\rm dir}
=
- \frac{1}{N_c}
\sum_{{\bf q} {\bf p}}
	\bar{V}_+
	{\rm Tr} \left(
		\langle \hat{\Theta}_{{\bf p}}^{\vphantom{\dagger}} \rangle
		\hat{\Theta}_{\bf q}^\dag
		+
		\langle \hat{\Theta}^\dag_{{\bf p}} \rangle
		\hat{\Theta}^{\vphantom{\dagger}}_{\bf q}
	\right).
\end{eqnarray}
Deriving this formula we replace the interaction function
$V_{+} ({\bf p} - {\bf q})$
by its average value at the Fermi surface. The constant
$\bar{V}_+$
is equal to
\begin{eqnarray}
\bar{V}_+ = {\frac14}\int_0^{2\pi} \frac{d\chi}{2\pi}
	(1 + \cos \chi) V_+ (k_{\rm F} \sqrt{2 - 2 \cos \chi}).
\end{eqnarray}
Likewise, the mean field form of the umklapp interaction is
\begin{eqnarray}
\hat{H}^{\rm MF}_{\rm u}
=
- \frac{1}{N_c}
\sum_{{\bf q} {\bf p}}
	\bar{V}_-
	{\rm Tr}\, \left(
		\langle \hat{\Theta}_{{\bf p}}^{\vphantom{\dagger}} \rangle
		\hat{\Theta}^{\vphantom{\dagger}}_{\bf q}
		+
		\langle \hat{\Theta}^\dag_{{\bf p}} \rangle
		\hat{\Theta}_{\bf q}^\dag
	\right),
\\
\bar{V}_- = {\frac14}\int_0^{2\pi} \frac{d\chi}{2\pi}
	(1 - \cos \chi) V_- (k_{\rm F} \sqrt{2 - 2 \cos \chi}).
\end{eqnarray}
As for
$\hat{H}_{\rm ex}$,
it does not contribute to the mean field Hamiltonian. Indeed, one can check
that non-zero expectation value
$\langle
	\hat{\gamma}^\dag_{{\bf q} 3 m}
	\hat{\gamma}^{\vphantom{\dag}}_{{\bf q} 2 m'}
\rangle$
in
Eq.~(\ref{eq::exchange})
is possible only at zero transferred momentum
${\bf k} = 0$.
Contributions with vanishing transferred momentum in
$\hat{H}_{\rm ex}^{\rm eff}$
vanish due to
\begin{eqnarray}
1- e^{ i (\phi_{\bf q} - \phi_{\bf q+k} )} =
1- e^{ i (\phi_{\bf q'} - \phi_{\bf q'-k} )} = 0
\end{eqnarray}
at
${\bf k} = 0$.

The resultant mean field Hamiltonian reads
\begin{eqnarray}
\label{eq::MF_Ham}
\hat{H}^{\rm MF} = \hat{H}^{\rm eff}_0 + \hat{H}^{\rm MF}_{\rm int},
\end{eqnarray}
where the mean field interaction is
\begin{eqnarray}
\label{eq::MF_interaction}
\hat{H}^{\rm MF}_{\rm int}
=
- \sum_{\bf q}
	{\rm Tr}\! \left(
		\hat{Q}^\dag \hat{\Theta}_{\bf q}^{\vphantom{\dagger}}
		+
		\hat{\Theta}_{\bf q}^\dag \hat{Q}^{\vphantom{\dagger}}
	\right).
\end{eqnarray}
In this expression the $4\times4$ matrix
$\hat{Q}$
is the order parameter
\begin{eqnarray}
\label{eq::0self-consist1}
\hat{Q}
=
\frac{1}{N_c}
\sum_{\bf p}\left(
	\bar{V}_+ \langle \hat{\Theta}_{\bf p}^{\vphantom{\dagger}} \rangle
	+
	\bar{V}_- \langle \hat{\Theta}_{\bf p}^\dag \rangle\right).
\end{eqnarray}

To derive a self-consistency equation, it is convenient to invert this
definition
\begin{eqnarray}
\label{eq::1self-consist1}
\frac{1}{N_c} \sum_{\bf p}
	\langle \hat{\Theta}_{\bf p} \rangle
=
\frac{1}{\bar{V}_+^2 - \bar{V}_-^2}
\left(
	\bar{V}_+ \hat{Q}^{\vphantom{\dagger}}
	-
	\bar{V}_- \hat{Q}^\dag
\right).
\end{eqnarray}
We prove in Appendix that for our mean field
Hamiltonian
$\hat{H}^{\rm MF}$
the symmetry-breaking average
$\langle \hat{\Theta}_{\bf p} \rangle$
satisfies
\begin{eqnarray}
\label{eq::hellm_feyn}
\frac{1}{N_c} \sum_{\bf p}
	\langle \hat{\Theta}_{\bf p} \rangle
=
\frac{1}{2 N_c}
\sum_{\bf k}
	\hat{Q} \left(
		\varepsilon_{\bf k}^2 + \hat{Q}^\dag\hat{Q}
	\right)^{-\frac{1}{2}}.
\end{eqnarray}
We consider only the undoped system, where the Fermi level is near Dirac
points and the energy spectrum has a rotational symmetry in the momentum
space. Therefore, it is reasonable to assume that the average
$\langle \Theta_{\mathbf p} \rangle$
is independent of the direction of
${\mathbf p}$.
Now, comparing the latter two equations and changing the summation over
momentum by integration over energy, we derive the self-consistency
equation in the form
\begin{eqnarray}
\label{eq::Qmatrix_self-consist}
(\bar{V}_+^2 - \bar{V}_-^2)
\hat{Q} h(\hat{Q}^\dag \hat{Q} )
=
\bar{V}_+ \hat{Q} - \bar{V}_- \hat{Q}^\dag,
\end{eqnarray}
in which the function $h$ is defined as
\begin{eqnarray}
\label{eq::h_def}
h(\hat{Q}^\dag \hat{Q} )
=
\frac{1}{2}
\int_0^{3t}\!\! \frac{\nu (\varepsilon) d \varepsilon}
	{\sqrt{( \varepsilon - t_0)^2 + \hat{Q}^\dag \hat{Q} }},
\end{eqnarray}
where
$\nu (\varepsilon) \approx \varepsilon/(\sqrt{3} \pi t^2)$
is the graphene density of states (per spin projection per valley) and integration is performed up to maximum electron energy in the AA-BLG~\cite{ourBLGreview2016}.

Now we briefly discuss certain mathematical points that must be settled before analysis of the derived equations.
Note a property of Eq.~\eqref{eq::Qmatrix_self-consist}: if $\hat{Q}_0$ is a solution of this equation, then $\hat{Z}^\dag\hat{Q}_0\hat{Z}$, where $\hat{Z}$ is a unitary matrix, is also a solution.
Observe that the radical of the matrix-valued polynomial in
Eqs.~(\ref{eq::hellm_feyn})
and~(\ref{eq::h_def})
is defined completely unambiguously. Indeed, the matrix
$\hat{Q}^\dag \hat{Q}$
is Hermitian positive semi-definite (that is, its eigenvalues are real and
non-negative). Such a matrix allows the following representation
\begin{eqnarray}
\label{eq::diag_QQ}
\hat{Q}^\dag \hat{Q} = \hat{V} \hat{D}^2 \hat{V}^\dag,
\
\hat{D} = {\rm diag}(d_1, \ldots, d_4),
\
d_i \geq 0,
\end{eqnarray}
where
$\hat{V}$
is a unitary matrix,
$\hat{V} \in {\rm U}(4)$.
Consequently,
\begin{eqnarray}
\label{eq::matrix_root}
\left( \varepsilon_{\bf k}^2 + \hat{Q}^\dag\hat{Q} \right)^{-\frac{1}{2}}
=
\hat{V}
\left( \varepsilon_{\bf k}^2 + \hat{D}^2 \right)^{-\frac{1}{2}}
\hat{V}^\dag.
\end{eqnarray}
An arbitrary function $f$ of a diagonal matrix is defined according to the convention
\begin{eqnarray}
f({\rm diag}\/[a_1,\ldots, a_n])
=
{\rm diag}\/[f(a_1),\ldots, f(a_n)],
\end{eqnarray}
assuming, of course, that
$f(a_i)$
are defined for all $i$'s. Applying this construction to
Eq.~(\ref{eq::matrix_root})
we can write
\begin{eqnarray}
\left( \varepsilon_{\bf k}^2 + \hat{D}^2 \right)^{-\frac{1}{2}}
\!\!\!=\!
{\rm diag}\!
\left[ \!
	\left(\varepsilon_{\bf k}^2 + d_1^2\right)^{-\frac{1}{2}}\!\!,
	\ldots,\!
	\left( \varepsilon_{\bf k}^2 + d_4^2 \right)^{-\frac{1}{2}}\!
\right]\!,
\end{eqnarray}
where the square root extraction is performed on real non-negative
quantities only.

The self-consistency
equation~(\ref{eq::Qmatrix_self-consist})
can be simplified, if we implement the singular-value decomposition on
$\hat{Q}$
and
$\hat{Q}^\dag$:
\begin{eqnarray}
\label{eq::def_UVD}
\hat{Q} = \hat{U} \hat{D} \hat{V}^\dag,
\quad
\hat{Q}^\dag = \hat{V} \hat{D} \hat{U}^\dag,
\end{eqnarray}
with matrices
$\hat{V}$
and
$\hat{D}$
being introduced in
Eq.~(\ref{eq::diag_QQ}).
As for
$\hat{U} \in$\,U(4),
it diagonalizes the product
$\hat{Q} \hat{Q}^\dag$,
that is,
$\hat{Q} \hat{Q}^\dag = \hat{U} \hat{D}^2 \hat{U}^\dag$.
We substitute Eqs.~\eqref{eq::def_UVD} in Eq.~\eqref{eq::Qmatrix_self-consist} and derive a diagonal form of the self-consistency equation
\begin{eqnarray}
\label{eq::self-consist_diag}
\left[ \bar{V}_+ - (\bar{V}_+^2 - \bar{V}_-^2) h(\hat{D}^2) \right] \hat{D}
=
\bar{V}_- \hat{W} \hat{D} \hat{W},
\end{eqnarray}
where
$\hat{W} = \hat{U}^\dag \hat{V} \in$\,U(4).

Let us assume that we solve the diagonalized self-consistency
equation~\eqref{eq::self-consist_diag},
that is, we obtain all possible pairs of the matrices
$\hat{D}_n$
and
$\hat{W}_n$
satisfying
Eq.~(\ref{eq::self-consist_diag}).
Then, we define
\begin{eqnarray}
\label{eq::D_times_W}
\hat{Q}_n = \hat{W}_n^\dag \hat{D}_n,
\quad
\hat{Q}^\dag_n = \hat{D}_n \hat{W}_n.
\end{eqnarray}
Direct substitution of Eqs.~\eqref{eq::D_times_W} in Eq.~(\ref{eq::Qmatrix_self-consist}) shows that $\hat{Q}_n$ is its solution. As it was stated above, all other solutions to the general self-consistency equation~(\ref{eq::Qmatrix_self-consist})
are unitary equivalent to matrices
$\hat{Q}_n$
defined by
Eq.~(\ref{eq::D_times_W}).

\section{Solutions of the self-consistency equation}
\label{sec::solution}

\subsection{The case of vanishing $\bar{V}_-$}
\label{subsec::zero_Vminus}

First, we consider a simplest case when
$\bar{V}_- = 0$.
This condition is not realistic but allows one to easily obtain an
analytical result. In so doing,
Eq.~(\ref{eq::self-consist_diag})
can be rewritten in the BCS-like form
\begin{eqnarray}
1 = \nu_0 \bar{V}_+ \ln \left(\frac{2 E^*}{\hat{D}} \right),
\end{eqnarray}
where
$\nu_0  \approx t_0/(\sqrt{3} \pi t^2)$
is the AA-BLG density of states at the Fermi level (per spin projection,
per valley, per single band), the energy scale is
$E^* = \sqrt{t_0(3t-t_0)}$.
The solution to this equation reads
\begin{eqnarray}
\label{eq::no_Vminus}
\hat{D} = \Delta_0 \mathbb{I}_4,
\quad
\text{where}
\quad
\Delta_0
=
2 E^* \exp \left[ - \frac{1}{\nu_0 \bar{V}_+ } \right],
\end{eqnarray}
and
$\mathbb{I}_4$
is the 4$\times$4 identity matrix. At the same time, when
$\bar{V}_-$ vanishes, the matrix $\hat{W}$
is not limited by the self-consistency equation. Lack of any
restrictions on $\hat{W}$ is the manifestation of the extended symmetry group $G_0$ of the model Hamiltonian in the case $\bar{V}_-=0$. As a result,
the order parameter matrix is
$\hat{Q} = \Delta_0 \hat{Y}$,
where
$\hat{Y}$
is an arbitrary unitary matrix. Note that four eigenvalues of
$\hat{Q}$
are equal to
$\Delta_0 \exp (i \alpha_{1, \ldots, 4})$,
where
$\alpha_i$
are arbitrary phases.

\subsection{Hermitian order parameters}
\label{subsec::hermitian_OP}

In a general case
$V_\pm \ne 0$,
the equation
system~(\ref{eq::self-consist_diag})
becomes much more complex. Now
$\hat{W}$
explicitly enters the self-consistency condition, drastically increasing
the number of unknown variables. In this paper we do not
attempt to find exhaustive solution to the problem. Instead, we will discuss three specific classes of the solutions of Eq.~(\ref{eq::self-consist_diag}) to illustrate the richness of the system under study.

It is natural to expect that in the ground state of our model the
single-electron gaps in the four fermionic sectors are identical to each
other. This situation can be represented by the ansatz
\begin{eqnarray}
\label{eq::iso_gap_ansatz}
\hat{D} = \Delta \mathbb{I}_4.
\end{eqnarray}
Substituting this into
Eq.~(\ref{eq::self-consist_diag}),
one establishes that
\begin{eqnarray}
\label{eq::W_ansatz}
\hat{W}^2 = a \mathbb{I}_4,
\quad
\text{where}
\quad
a = \pm 1.
\end{eqnarray}
Let us consider first
$a = +1$.
Then
\begin{eqnarray}
(\bar{V}_+ + \bar{V}_-) h(\Delta^2) = 1,
\quad
\hat{W}^2 = \mathbb{I}_4.
\end{eqnarray}
The solution for $\Delta$ is
\begin{eqnarray}
\label{eq::Delta+}
\Delta = 2E^* \exp \left[ -\frac{1}{\nu_0 (\bar{V}_+ + \bar{V}_-)} \right].
\end{eqnarray}
A unitary matrix $\hat{W}$, whose square is
$\mathbb{I}_4$,
can be expressed as
$\hat{W} = \hat{S} \hat{\mathbf{\Sigma}} \hat{S}^\dag$,
where
$\hat{S}$
is a unitary matrix, and
$\hat{\mathbf{\Sigma}}$
is a diagonal matrix whose elements are $\pm 1$.
In such a case, the order parameter matrix is Hermitian and satisfies
\begin{eqnarray}
\label{eq::hermitian_Q_represent}
\hat{Q} = \hat{Q}^\dag = \Delta \hat{Z} \hat{\mathbf{\Sigma}} \hat{Z}^\dag,
\end{eqnarray}
where
$\hat{Z} \in$\, SU(4).
In other words, the order parameter is equal to
$\Delta \hat{\mathbf{\Sigma}}$
up to a unitary transformation.

To classify all types of
$\hat{Q}$
consistent with
Eq.~(\ref{eq::hermitian_Q_represent}),
we split all possible solutions into three
topologically distinct classes labeled by
$\alpha \in \left\{ {\rm I, II, III} \right\}$.
These classes are defined by the structure of
$\hat{\mathbf{\Sigma}}$
\begin{eqnarray}
\label{eq::classI}
\text{Class I:}&
\hat{\mathbf{\Sigma}}_{\rm I}^\kappa = {\rm diag}\,(\kappa, \kappa, \kappa, \kappa),
&
\kappa = \pm 1,
\\
\label{eq::classII}
\text{Class II:}&
\hat{\mathbf{\Sigma}}_{\rm II} = {\rm diag}\,(1,1,-1,-1),
\\
\label{eq::classIII}
\text{Class III:}&
\hat{\mathbf \Sigma}_{\rm III}^\kappa= {\rm diag}\,(\kappa, \kappa, \kappa, -\kappa),
&
\kappa = \pm 1.
\end{eqnarray}
Any
$\hat{Q}$
satisfying
condition~(\ref{eq::hermitian_Q_represent})
belongs to one and only one class among these three.

\textit{Class-I} order parameter is a charge-density wave state (CDW). To
demonstrate this let us first observe that
$\hat{\mathbf{\Sigma}}_{\rm I}^\kappa \propto \mathbb{I}_4$ and
the order parameter is independent of
$\hat{Z}$. Thus, it always satisfies
$\hat{Q} = \hat{Q}^\dag = \kappa\Delta \mathbb{I}_4$ and
$\hat{Q}$
is diagonal in the multi-index
$m=(\xi, \sigma)$
space. Consequently, when calculating local occupation numbers
$n_{l a} = \sum_\sigma
\langle
	\hat{d}^\dag_{\mathbf{n}la\sigma}
	\hat{d}^{\vphantom{\dagger}}_{\mathbf{n}la\sigma}
\rangle$,
only ``diagonal" symmetry-breaking averages
$\langle
	\hat{\gamma}_{\mathbf{q} 3 \xi \sigma }^\dag
	 \hat{\gamma}_{\mathbf{q}2 \xi \sigma}^{\vphantom{\dagger}}
\rangle$
are non-zero. Keeping this in mind and using Eqs.~\eqref{eq::GAMMA_valley} and~\eqref{real_space_op}, we derive
\begin{eqnarray}
n_{l a} = 1 + \sum_m \delta n_{l a m},
\end{eqnarray}
where the anomalous contributions to
$n_{la}$
are
\begin{eqnarray}\label{delta_n_I}
\delta n_{l a m}
=
\frac{(-1)^{a+l}}{2 N_c}
\sum_{\bf q}
	{\rm Re}\,
	\langle
		\hat{\gamma}_{\mathbf{q} 3 m}^\dag
		 \hat{\gamma}_{\mathbf{q}2 m}^{\vphantom{\dagger}}
	\rangle.
\end{eqnarray}
These two relations can be compactly written as
\begin{eqnarray}
n_{l a}
=
1 +
\frac{(-1)^{a+l}}{2 N_c}
{\rm Tr}\, \sum_{\bf q}
	{\rm Re}\, \langle \hat{\Theta}_{\bf q} \rangle.
\end{eqnarray}
Equation~(\ref{eq::1self-consist1}),
allows us to connect
$\sum_{\bf q} \langle \hat{\Theta}_{\bf q} \rangle$
with
$\hat{Q}$, and
finally we obtain
\begin{eqnarray}
\label{eq::cdw_charge_distrib}
n_{l a} = 1 + \kappa (-1)^{a+l} \frac{2\Delta}{\bar{V}_+ + \bar{V}_-}.
\end{eqnarray}
Thus, the electric charge distribution is inhomogeneous within a single unit cell,
the strength of the inhomogeneity is proportional to $\Delta$. This
corresponds to a commensurate CDW state.

The commensurate order parameter does not violate the translation symmetry
of the underlying honeycomb lattice. However, the symmetry between the
sublattices is broken, the same is true for the symmetry between the
layers. Analyzing
Eq.~(\ref{eq::cdw_charge_distrib})
we note that switching layers or sublattices is equivalent to switching the
sign of $\kappa$. Thus, the inversion of $\kappa$ is associated with the
$\mathbb{Z}_2$ subgroup of the symmetry group $G$.

\textit{Class~II} contains six mutually unitary-equivalent diagonal matrices
$\hat{Q}$.
These are
$\pm \hat{\mathbf{\Sigma}}_{\rm II}$
and
\begin{eqnarray}
\pm {\rm diag}\, (1, -1, 1, -1)
\quad
\text{and}
\quad
\pm {\rm diag}\, (1, -1, -1, 1).
\end{eqnarray}
Any such matrix can be obtained from
$\hat{\mathbf{\Sigma}}_{\rm II}$
by a suitable permutation of its diagonal elements.

If the matrix order parameter
$\hat{Q}$ 
of the class II is diagonal, then, we derive similar to 
Eq.~\eqref{delta_n_I}
\begin{eqnarray}
|\delta n_{lam} | = \frac{\Delta}{2(\bar{V}_+ + \bar{V}_-)}.
\end{eqnarray}
For fixed $l$ and $a$, two of
$\delta n_{lam}$
are positive, while two are negative. Thus, unlike the class-I CDW, electronic phases
of class~II have a homogeneous charge distribution within a unit cell since
$\sum_m \delta n_{lam} \equiv 0$
for any $l$ and $a$.
Depending on which
$\delta n_{lam}$'s
are positive, and which are negative, three distinct types
of the order can be distinguished: spin-density wave (SDW), valley-density wave (VDW), and
spin-valley-density wave (SVDW). For example, the choice
\begin{eqnarray}
\delta n_{la \xi\sigma} = \pm \sigma (-1)^{a+l} |\delta n_{la \xi\sigma}|
\end{eqnarray}
corresponds to the SDW phase in which the expectation value of the spin operator
${\hat{S}}^z_{la}$
is finite
\begin{eqnarray}
\langle {\hat{S}}^z_{la} \rangle
=
\sum_{\xi \sigma} \sigma \delta n_{la \xi \sigma}
=
\pm (-1)^{a+l} \frac{2\Delta}{\bar{V}_+ + \bar{V}_-}.
\end{eqnarray}
When
$\delta n_{la \xi \sigma } \propto (-1)^\xi$,
the system is in the VDW phase, with finite staggered valley polarization
\begin{eqnarray}
\langle {\hat{S}}^v_{la} \rangle
=
\sum_{\xi \sigma} (-1)^\xi \delta n_{la \xi \sigma}
=
\pm (-1)^{a+l} \frac{2 \Delta}{\bar{V}_+ + \bar{V}_-}.
\end{eqnarray}
Finally, the SVDW order corresponds to
$\delta n_{la \xi \sigma } \propto (-1)^\xi \sigma$.
This phase has finite staggered spin-valley polarization
$\langle {\hat{S}}^{sv}_{la} \rangle
=
\sum_{\xi \sigma} (-1)^\xi \sigma \delta n_{la \xi \sigma}$.

While the class-I order parameter is always the same for any
$\hat{Z}$,
the class-II matrix
$\hat{Q}$
changes when
$\hat{Z}$
in
Eq.~(\ref{eq::hermitian_Q_represent})
is changed. For example, a suitably chosen
$\hat{Z}$
connects all three class-II phases to each other. 

Furthermore, in class-II, matrix
$\hat{Q}$
does not have to be diagonal: for a generic choice of
$\hat{Z}$,
non-zero elements connecting different spin projections and different
valleys are possible. Recall that, if an order parameter is non-diagonal in
spin indices, it represents spin polarization deviating from the $z$-axis.
Additionally,
$\hat{Q}$
can be non-diagonal in valley indices. In real space, these inter-valley
matrix elements correspond to spatially oscillating contributions, with
${\bf K}_1 - {\bf K}_2$ 
being their wave vector. When such an inter-valley coherence is realized,
the elementary translation vectors are tripled in length.

In \textit{class-III}, the diagonal order parameter
$\hat{Q} \propto {\hat{\mathbf{\Sigma}}}^\kappa_{\rm III}$
represents a state with finite polarizations with respect to all four types
of density waves (CDW, SDW, VDW, and SVDW). Unitary matrix
$\hat{Z}$
affects the SDW, VDW, and SVDW polarizations. The sign of the CDW order
parameter can be changed only by the inversion of $\kappa$. The
inter-valley coherence is also possible in this class.

Finally, we would like to note that in the considered case the order
parameter eigenvalues are always real and equal to $\pm \Delta$, since
any Hermitian solution
$\hat{Q}$
discussed above is unitary-equivalent to
$\Delta \hat{\mathbf{\Sigma}}$.

\subsection{Anti-Hermitian order parameters}
\label{subsec::anti_hermitian_OP}

The order parameter matrix 
$\hat{Q}$ 
does not always have to be Hermitian. To illustrate this point, let us
choose 
$a=-1$
in 
Eq.~\eqref{eq::W_ansatz}. 
In this case we have
\begin{eqnarray}
\label{eq::imag_W}
\hat{W}=i\hat{S}\hat{\mathbf{\Sigma}}\hat{S}^{\dag},
\end{eqnarray}
where the structures of the matrices
$\hat{S}$
and 
$\hat{\mathbf{\Sigma}}$
are defined in the previous subsection. Assuming that
$\hat{D} \propto \mathbb{I}_4$
as in
Eq.~(\ref{eq::iso_gap_ansatz}),
we substitute this
$\hat{W}$
in
Eq.~(\ref{eq::self-consist_diag}) and obtain
four identical mutually decoupled
equations. Solving these equations, we derive
\begin{eqnarray}
\label{eq::Delta-}
\hat{D} = \Tilde \Delta \mathbb{I}_4,
\quad
\text{where}
\quad
\Tilde \Delta
=
2 E^* \exp \left[ - \frac{1}{\nu_0 (\bar{V}_+ - \bar{V}_-)} \right].
\quad
\end{eqnarray}
The sign before $\bar{V}_-$
is the only difference between $\Delta$ and
$\Tilde \Delta$.
The order parameter matrix now reads
\begin{eqnarray}
\label{eq::antiH_Q}
\hat{Q} = -\hat{Q}^\dag
=
-i\tilde{\Delta}\hat{Z}\hat{\mathbf{\Sigma}} \hat{Z}^{\dag},
\end{eqnarray}
where 
$\hat{Z}\in$\, SU(4).
Thus, the matrix
$\hat{Q}$
is anti-Hermitian.

Following the logic of
Sec.~\ref{subsec::hermitian_OP},
let us consider a special case of     
$\hat{\Sigma}=\chi\mathbb{I}_4$, 
where 
$\chi=\pm1$.
From
Eqs.~(\ref{eq::1self-consist1})
and~(\ref{eq::antiH_Q})
we obtain that
\begin{eqnarray}\label{aver_Q_antiHer}
\frac{1}{N_c} \sum_{\bf p}
	\langle \hat{\Theta}_{\bf p}^{\vphantom{\dagger}} \rangle
=
- \frac{1}{N_c} \sum_{\bf p}
	\langle \hat{\Theta}_{\bf p}^\dag \rangle
=
- \frac{i \chi \Tilde \Delta}{\bar{V}_+ - \bar{V}_-}.
\end{eqnarray}
Since
$\frac{1}{N_c} \sum_{\bf p}
\langle \hat{\Theta}_{\bf p}^{\vphantom{\dagger}} \rangle$
is purely imaginary, then
$\delta n_{lam} = 0$.
On the other hand, ``inter-layer" current operator
\begin{eqnarray}
\hat{I}^\perp_{{\bf n} a}
=
i \sum_\sigma
	\hat{d}^\dag_{\mathbf{n}1a\sigma}
	\hat{d}^{\vphantom{\dagger}}_{\mathbf{n}2a\sigma}
+ {\rm H.c.}
\end{eqnarray}
has a finite expectation value. Let us prove it.

The expectation value
$\langle \hat{I}^\perp_{{\bf n} a} \rangle$
can be presented as a sum over the multi-index $m$
\begin{eqnarray}
\langle  \hat{I}^\perp_{{\bf n} a} \rangle
=
\sum_m \langle  \hat{I}^\perp_{a m} \rangle,
\end{eqnarray}
where we assumed that the average
$\langle  \hat{I}^\perp_{{\bf n} a} \rangle$
is independent of
${\bf n}$
due to ground-state translation invariance. The partial current
$\langle \hat{I}^\perp_{ a m} \rangle$
is equal to
\begin{eqnarray}
\langle \hat{I}^\perp_{ a m} \rangle
=
\frac{1}{2N_c} \sum_{{\bf k} }
	{\rm Im}\/
	\Big<\!\!
		\left[
			\hat{\gamma}^\dag_{{\bf k} 2 m}\!\!
			+
			(-1)^{a} \hat{\gamma}^\dag_{{\bf k} 3 m}
		\right]
	\times
\\
\nonumber
		\left[
			\hat{\gamma}_{{\bf k} 2 m}^{\vphantom{\dag}} \!\!
			-
			(-1)^{a}
			\hat{\gamma}_{{\bf k} 3 m}^{\vphantom{\dag}}
		\right]\!\!
	\Big>.
\end{eqnarray}
Combining the latter equation and
Eq.~(\ref{aver_Q_antiHer}),
one derives
\begin{eqnarray}
\langle \hat{I}^\perp_{ a m} \rangle
=
\frac{(-1)^{a}}{N_c} \sum_{{\bf k} }
	{\rm Im}\/
	\langle
		\hat{\gamma}^\dag_{{\bf k} 3 m}
		\hat{\gamma}^{\vphantom{\dag}}_{{\bf k} 2 m}
	\rangle
= \frac{\chi (-1)^a \Tilde \Delta}{\bar{V}_+ - \bar{V}_-}.
\end{eqnarray}
Consequently
\begin{eqnarray}
\langle \hat{I}^\perp_{{\bf n} a} \rangle
=
\chi (-1)^a \frac{4 \Tilde \Delta}{\bar{V}_+ - \bar{V}_-}.
\end{eqnarray}
We see that such a state is characterized by spontaneously generated
inter-layer currents. Factor
$(-1)^a$
in the expression for
$\langle  \hat{I}^\perp_{{\bf n} a} \rangle$
indicates that the flow along the inter-layer links on sublattice $A$
exactly cancels the flow along the inter-layer links on sublattice $B$.
Therefore, the overall inter-layer charge flow is zero, as it must be in an
eigenstate. Note also, a detailed distribution of the current flow is
impossible to calculate within the current formalism since we neglect the
existence of the bands~1 and~4.

One can adopt the reasoning of
subsection~\ref{subsec::hermitian_OP}
and introduce a topological classification of the anti-Hermitian order
parameters: there are three distinct classes, whose order parameters are
identical, up to multiplication on complex unity $i$, to the order
parameters in classes~I, II, and III discussed above. Instead of the
spontaneous local densities, the ordered states corresponding to the
anti-Hermitian
$\hat{Q}$
are characterized by spontaneous inter-layer currents. Depending on the
topological class, these currents may carry charge, spin, valley,
spin-valley quanta, or combinations of the above.

\subsection{Non-Hermitian non-anti-Hermitian order parameters}
\label{subsec::non_hermit}

\begin{table}[ht]
\centering
\begin{tabular}{||c|c|c||}
\hline
1 real eigenvalue &
2 real eigenvalues &
3 real eigenvalues \\
$(\pm 1, \pm i, \pm i, \pm i)$&
$(\pm 1, \pm 1, \pm i, \pm i)$ &
$(\pm 1, \pm 1, \pm 1, \pm i)$ \\
\hline
\hline
$(\kappa, i\chi, i\chi, i\chi)$&
$(\kappa, \kappa, i\chi, i\chi)$&
$(\kappa, \kappa, \kappa, i\chi)$\\
\hline
$(\kappa, i\chi, i, -i)$&
$(\kappa, \kappa, i, -i)$&
$(\kappa, 1, -1, i\chi)$\\
\hline
& $(1, -1, i\chi, i\chi)$& \\
\hline
& $(1, -1, i, -i)$& \\
\hline
\hline
\end{tabular}
\caption{Classification of non-Hermitian non-anti-Hermitian order
parameters. All matrices
$\hat{Q}^\dag$
described by
Eq.~(\ref{eq::non_herm_OP})
can be split into three types according to the number of real eigenvalues.
Each column of the table represents one of of these types. We use the following notations:
$\pm 1$
stands for a real eigenvalue,
$\pm i$
stands for  an imaginary eigenvalue, and the binary indices
$\kappa = \pm 1$
and
$\chi = \pm 1$
are the same as in
subsections~\ref{subsec::hermitian_OP}
and~\ref{subsec::anti_hermitian_OP}.
Within each type, additional sub-types can be defined, according to the number
of minus signs in front of real and imaginary eigenvalues (up to a
permutation of eigenvalues). Any diagonal order parameter matrix set by
Eq.~(\ref{eq::non_herm_OP})
belongs to one and only one sub-type defined in this table.
\label{table::classification}
}
\end{table}

In this subsection we demonstrate that the order parameter
$\hat{Q}$
satisfying the self-consistency
equation~(\ref{eq::Qmatrix_self-consist})
may be neither Hermitian, nor anti-Hermitian. To prove this point, let
us consider the matrix set
$\frak{M}$
that contains
$4^4$
diagonal matrices of the following structure
\begin{eqnarray}
\label{eq::most_gen_W}
\hat{W} = {\rm diag}\, (w_1,\ldots, w_4),
\quad
\text{where}
\quad
w_m = \pm 1, \pm i.
\end{eqnarray}
Any diagonal matrix that satisfies 
condition~(\ref{eq::W_ansatz})
unavoidably satisfies 
Eq.~(\ref{eq::most_gen_W}). 
Therefore, all matrices 
$\hat{W}$
discussed in
subsections~\ref{subsec::hermitian_OP}
and~\ref{subsec::anti_hermitian_OP}
belong to the set 
$\frak{M}$.
The inverse statement is obviously not true.

The matrices that belong to
$\frak{M}$
but violate
condition~(\ref{eq::W_ansatz})
can be described in terms of their diagonal elements (up to a permutation)
as follows
\begin{eqnarray}
w_1 = \pm 1,
\quad
w_2 = \pm 1, \pm i,
\quad
w_3 = \pm 1, \pm i,
\quad
w_4 = \pm i.
\quad
\end{eqnarray}
For matrices
$\hat{W}$
of this type the self-consistency condition splits into four decoupled
equations. However, not all of these equations are identical. It is easy to
demonstrate that the diagonal elements of
$\hat{D}$
satisfy
\begin{eqnarray}
d_m
=
\begin{cases}
	\Delta, & \text{if}\quad w_m = \pm 1, \\
	\Tilde \Delta, & \text{if}\quad w_m = \pm i. \\
\end{cases}
\end{eqnarray}
Diagonal order parameter
$\hat{Q}^\dag$
equals
\begin{eqnarray}
\label{eq::non_herm_OP}
\hat{Q}^\dag = {\rm diag}\/ (w_1 d_1, \ldots, w_4 d_4).
\end{eqnarray}
Non-diagonal
$\hat{Q}^\dag$
can be obtained by application of a unitary transformation. We clearly see
that all such order parameters are neither Hermitian, nor anti-Hermitian.

The order
parameter~(\ref{eq::non_herm_OP})
represents a state that includes the features of both the Hermitian and the anti-Hermitian order-parameter states. If
$w_m$ is real, the non-vanishing symmetry-breaking observable for this
multi-index $m$ is the local density
$\langle \delta n_{lam} \rangle \ne 0$,
as in
subsection~\ref{subsec::hermitian_OP},
otherwise, it is the inter-layer current
$\langle I^\perp_{a m} \rangle \ne 0$,
as in
subsection~\ref{subsec::anti_hermitian_OP}.
Table~\ref{table::classification}
presents a classification scheme for these ordered phases.

\section{Discussion}
\label{sec::Discussion}
Due to peculiar features of the honeycomb lattice, the single-electron
dispersion in graphene and graphene-based systems is characterized by an
additional quantum number, valley index. Although, in many respects, the
valley index differs from the spin projection, it is possible to formulate
a theory that incorporates these two quantum number on an equal footing.
Our paper presents such a theory for the specific case of AA-BLG.

An SU(4)-symmetric [SU(4)S] theory of a graphene-based system
cannot serve as an ultimate model describing electronic properties in
detail. Yet, it is a helpful theoretical tool. Let us recall
that the presence of the valley degeneracy in graphene-based materials
opens new possibilities for electron-electron scattering and electron
low-temperature ordering. In such a situation, an accurate ``bookkeeping"
of all scattering and ordering channels may be quite challenging. A study
of an SU(4)S-model should be viewed as a physically-motivated approach aiming
at developing a concise classification scheme of the ordered states in
graphene-based materials.

The discussion presented above attests both to difficulties that one faces when
trying to itemize all allowed ordered phases in AA-BLG, and to usefulness
of an SU(4)S model for such an endeavor. Our main result here is the
derivation of the self-consistency
equation~(\ref{eq::self-consist_diag})
and the list (possibly, incomplete) of ordered phases satisfying this
equation. One should appreciate the length of this list, as well as the fact
that all these dissimilar many-body states have been identified within a
single unifying approach, as solutions to
Eq.~(\ref{eq::self-consist_diag}).

At the same time, the proposed method suffers from several limitations that require
additional research. One must remember that a single state
with lowest energy inevitably becomes the true ground state. Trying to use
the SU(4)S model to determine which state is the ground state, we discover
that quite dissimilar phases aggregate into broad multiplicities, with all
phases in a multiplicity being degenerate and connected to each other by
suitable SU(4) Bogolyubov transformations. For example, class~II of the
Hermitian ordered phases unites SDW, VDW, and SVDW into a single group of
degenerate states. This ``blindness" of the classification is a consequence
of consideration of the spin and valley indices in equal
footing.

Clearly, a more realistic theory must distinguish the valley quantum, of
purely orbital origin, and spin, a consequence of the relativistic
Dirac-equation physics. In a general situation one expects that non-SU(4)S
terms in the Hamiltonian destroy the spin-valley symmetry, and lift
multiple degeneracies of the SU(4)S model. In this respect, we already
identified the back-scattering interaction as a non-SU(4)S term. Other
possible non-SU(4)S contributions may emerge when electron-lattice
coupling is taken into account. The short-range interaction is also
incompatible with the SU(4)S. Additionally, external influences (electric
and/or magnetic fields, substrate choice, deformations, etc.) engineered
for a specific purpose (e.g., stabilizing a specific type of order
parameter) must be considered as well.

Under such circumstances, we expect that the true ground state will be
chosen from the list as a result of the interplay of various non-universal
and, possibly, sample-specific factors. Situations of this sort, when
multiple states compete against each other to become the true ground state,
are known to appear in doped Hubbard
model~\cite{kokanova2021prb, corboz_rice2014tJcompetition,
white_hubb_stripes2017numerics},
and models with
nesting~\cite{sboychakov_FraM2021prb_lett,PrbROur,
Sboychakov_PRB2013_PS_AAgraph,
our_hmet_prl2017,
separation_pnictides2013,
separation_SDW2013,
tokatly1992}.
Unlike these, for our SU(4)S model such competition occurs already at zero
doping. However, the aspect common for both types of models is the
importance of numerous non-universal contributions affecting the final outcome
of the competition~\cite{kokanova2021prb}.

Two additional questions for the future research are (i)~the completeness of
the ordered phase list and (ii)~the stability of the phases on that list.
In connection to (i) we must say that within our formalism this problem
becomes a purely mathematical task of exhausting all possible solutions to
Eq.~(\ref{eq::self-consist_diag}).
Currently, we do not know if ordered states other than those discussed in
Sec.~\ref{sec::solution}
can be identified.

As for (ii), we want to emphasize that a stability study of an ordered
state may be very complicated, and any self-consistency equation is
insufficient to establish stability or metastability of its solutions. For
a mean field theory, like ours, one can compare mean field energies for
various states. For example, it is easy to show that, for positive
$\bar{V}_+$,
the Hermitian order parameter state has lower energy than the
anti-Hermitian when
$\bar{V}_->0$.
For negative
$\bar{V}_-$,
the anti-Hermitian order parameter states have lower mean field energy. For
any sign of
$\bar{V}_-$,
the order
parameter~(\ref{eq::non_herm_OP})
has the energy that is in between the two. 

The latter argumentation, however, ignores the issue of the non-mean-field
fluctuations. They are important for two reasons: the fluctuation-induced
contributions to the energy can potentially lift the degeneracies between
different 
multiplicities~\cite{Nandkishore2010b},
and the fluctuations can completely destroy the order through the
Mermin–Wagner–Hohenberg mechanism. These two problems are the two
sides to the same looming question: the reliability of the mean field
theory. While, at present, it is impossible to address this question in
full generality, a good measure of theoretical understanding is already
available. For AA-BLG and other two-dimensional systems, it is expected
that the mean field theory is valid at zero temperature, at least
qualitatively. Zero-temperature fluctuations corrections to the mean field
energy may be of the same order as the mean field energy
itself~\cite{Nandkishore2010b}.
This does present a certain theoretical 
difficulty~\cite{mean_field_forDW2018schmalian}.
Fortunately, these corrections often can be
interpreted~\cite{kokanova2021prb, mean_field_forDW2018schmalian}
as (weak) renormalizations to the interaction constants, which allows one
to preserve the general mean field structure of the theory.

As for Mermin–Wagner–Hohenberg mechanism, it is well-recognized that any
non-Ising order in a two-dimensional system cannot endure finite
temperatures. Yet, the destroyed order does not disappear completely, but
rather survive in the form of short-range correlations, which gradually
vanish through a crossover. For a specific case of the SDW in AA-BLG,
qualitative theory of this crossover was discussed in
Ref.~\onlinecite{Sboychakov_PRB2013_PS_AAgraph}.

As we noted in
Sec.~\ref{intro},
our analysis of the order parameter symmetries in the AA-BLG is an
extension of the approach proposed by R.~Nandkishore and L.~Levitov in
Ref.~\onlinecite{Nandkishore2010b} for
AB-BLG. However, the authors of
Ref.~\onlinecite{Nandkishore2010b}
limited themselves by the study of Hermitian order parameters only. We
guess that including in the consideration non-Hermitian orders in the
AB-BLG is of interest and can be performed using the present approach. Of a
particular interest is the study of the order parameters in twisted bilayer
graphene. At low twist angles (and at magical angles, in particular), this
system can be considered as a periodic arrangement of regions with AA and
AB stacking.  Thus, the analysis of the possible symmetries of the order
parameters in ``aligned'' bilayers (AB-BLG and AA-BLG) can be considered as
a first step for study of the twisted graphene systems.

Finally, we want to mention that the proposed approach may be useful for
classification of superconducting order parameters in graphene-based
systems. However, in this case the consideration requires a significant
modification. In particular, we should take into account effects of doping.
So, the analyzes of the superconducting orders is beyond the scope of
present work and is the topic of further studies.

To conclude, we present a SU(4)-invariant model for the AA-BLG and investigate
this model within the mean field approximation. The derived matrix
self-consistency equation demonstrates rich diversity of solutions, every
solution representing a stationary ordered many-body phase of our model.
This wealth of the ordered states with close energies indicates that in
the AA-BLG several ordered phases compete against each other to become the true
ground state. Symmetry-based classification of the discussed phases is
developed.

\section*{Acknowledgment}
This research was funded by Russian Science Foundation Grant
No.~22-22-00464,
\url{https://rscf.ru/en/project/22-22-00464/}.

\appendix

\section{Calculation of the symmetry-breaking average}
\label{sec::Hellmann_Feyn}

To derive the self-consistency relation in the form of Eq.~\eqref{eq::hellm_feyn} we need to express
$N_c^{-1} \sum_{\bf q} \langle \hat{\Theta}_{\bf q} \rangle$
in terms of
$\hat{Q}$.
A convenient approach to address this task is to use the Hellmann-Feynman
formula
\begin{eqnarray}
\label{eq::hellmann_feynman}
\Bigl< \frac{\partial \hat H }{ \partial \lambda } \Bigr>
=
\frac{\partial E_0 }{ \partial \lambda},
\end{eqnarray}
where
$\hat{H}$
is a Hamiltonian dependent on some parameter $\lambda$,
and
$E_0 = E_0 (\lambda)$
is the ground state energy of
$\hat{H} (\lambda)$.

To adopt the latter formula to our mean field approximation, we need to look
for extrema of
$E^{\rm MF} = E^{\rm MF} (\hat{Q}, \hat{Q}^\dag)$
over
$\hat{Q}$.
Since
$\hat{Q}$
is a matrix, it is useful to introduce the differentiation over a matrix.
Namely, the derivative
$\frac{\partial f(\hat{X})}{\partial \hat{X}}$
is a matrix
$\hat{Y} = \frac{\partial f}{\partial \hat{X}}$
whose elements
$y_{ss'}$
are equal to
\begin{eqnarray}
\label{eq::matrix_diff_defin}
y_{ss'} = \frac{\partial f}{\partial x_{s's}},
\end{eqnarray}
where
$x_{ss'}$
are elements of
$\hat X$.
This definition implies that
\begin{eqnarray}
\label{eq::trace_diff}
\frac{\partial }{\partial \hat{X}} {\rm Tr}\/ (\hat{A} \hat{X}) = \hat{A},
\end{eqnarray}
provided that
$\hat{A}$
itself is independent of
$\hat{X}$.
Using these notations and the theorem~(\ref{eq::hellmann_feynman}),
we obtain for our mean field 
Hamiltonian~(\ref{eq::MF_Ham})
\begin{eqnarray}
\label{eq::Hell_Feyn_matrix1}
\frac{1}{N_c}\sum_{\bf q} \langle \hat{\Theta}_{\bf q} \rangle
=
- \frac{\partial E^{\rm MF}}{\partial \hat{Q}^\dag},
\\
\label{eq::Hell_Feyn_matrix2}
\frac{1}{N_c}\sum_{\bf q} \langle \hat{\Theta}_{\bf q}^\dag \rangle
=
- \frac{\partial E^{\rm MF}}{\partial \hat{Q}},
\end{eqnarray}
where
$E^{\rm MF}$
is the ground state energy (per unit cell) for
$\hat{H}^{\rm MF}$,
see
Eq.~(\ref{eq::MF_interaction}).

To calculate
$E^{\rm MF}$
it is convenient to write
$\hat{H}^{\rm MF}$, Eq.~\eqref{eq::MF_Ham},
as follows
\begin{eqnarray}
\label{eq::H_MF}
\hat{H}^{\rm MF}
=
\sum_{\bf q}
	\Phi^\dag_{\bf q}
	\hat{\cal H}_{\bf q}
	\Phi^{\vphantom{\dagger}}_{\bf q}.
\end{eqnarray}
In this formula the eight-component vector
$\Phi^\dag_{\bf q}$
equals
\begin{eqnarray}
\Phi^\dag_{\bf q} = (\Psi^\dag_{{\bf q}2}, \Psi^\dag_{{\bf q}3}),
\end{eqnarray}
where band-specific vectors
$\Psi^\dag_{{\bf q} s}$
($s=2,3$)
are introduced according to
\begin{eqnarray}
\Psi^\dag_{{\bf q}s}
=
(\hat{\gamma}_{{\bf q} s \uparrow {\bf K}_1}^\dag,
\hat{\gamma}_{{\bf q} s \downarrow {\bf K}_1}^\dag,
\hat{\gamma}_{{\bf q} s \uparrow {\bf K}_2}^\dag,
\hat{\gamma}_{{\bf q} s \downarrow {\bf K}_2}^\dag).
\end{eqnarray}
Symbol
$\hat{\cal H}_{\bf q}$
in
Eq.~(\ref{eq::H_MF})
is the $8\times 8$ matrix defined as
\begin{eqnarray}
\hat{\cal H}_{\bf q}
=
\left( \begin{matrix}
	\varepsilon_{\bf q}  \mathbb{I}_4 & - \hat{Q}^{\rm T} \\
	- \hat{Q}^* & - \varepsilon_{\bf q} \mathbb{I}_4 \\
	\end{matrix}
\right),
\end{eqnarray}
where $\varepsilon_{\bf q} = v_{\rm F} |{\bf q}| - t_0$.

The energy
$E^{\rm MF}$
then equals
\begin{eqnarray}
E^{\rm MF}
=
\frac{1}{N_c} \sum_{\bf q} \sum_{n=1,\ldots, 8}
	\mathcal{E}_{\bf q}^{(n)}
	\vartheta \left(-\mathcal{E}_{\bf q}^{(n)} \right),
\end{eqnarray}
where
$\vartheta (x)$
is the step-function, and
$\mathcal{E}_{\bf q}^{(n)}$
is the $n$'th eigenvalue of
$\hat{\cal H}_{\bf q}$.
Thus, we need to calculate the eigenvalues of
$\hat{\cal H}_{\bf q}$.
It is more convenient to replace the matrix
$\hat{\cal H}_{\bf q}$
with its complex conjugate
$\hat{\cal H}_{\bf q}^*$,
which, however, has the same set of
$\mathcal{E}_{\bf q}^{(n)}$.
The eigenvalue/eigenvector relation for
$\hat{\cal H}_{\bf q}^*$
reads
\begin{eqnarray}
\left(
\begin{matrix}
	\varepsilon_{\bf q} & -\hat{Q}^\dag \\
	- \hat{Q} & - \varepsilon_{\bf q} \\
\end{matrix}
\right)
\left(
	\begin{matrix}
		\phi_1\\
		\phi_2\\
	\end{matrix}
\right)
=
\mathcal{E}_{\bf q}
\left(
	\begin{matrix}
		\phi_1\\
		\phi_2\\
	\end{matrix}
\right).
\end{eqnarray}
Excluding
$\phi_2$
from this equation, we derive
\begin{eqnarray}
\left( \mathcal{E}_{\bf q}^2 - \varepsilon_{\bf q}^2 \right) \phi_1
=
\hat{Q}^\dag \hat{Q} \phi_1.
\end{eqnarray}
Thus, the eigenenergies of
$\hat{\cal H}_{\bf q}$
are
\begin{eqnarray}
\mathcal{E}_{\bf q}^{(n)}
=
\pm \sqrt{ \varepsilon_{\bf q}^2 + d_i^2 },
\end{eqnarray}
where
$d_i^2$,
$i = 1, \ldots, 4$,
are the eigenvalues of the positive semi-definite 4$\times$4 matrix
$\hat{Q}^\dag\hat{Q}$.
Since the trace of a matrix is invariant under unitary transformations, we derive
\begin{eqnarray}
\label{eq::E_MF_QQ}
E^{\rm MF}
=
- \frac{1}{N_c} \sum_{\bf q}
	{\rm Tr}\!
	\left(
		\varepsilon_{\bf q}^2 + \hat{Q}^\dag \hat{Q}
	\right)^{\frac{1}{2}}.
\end{eqnarray}
The final step is to find the derivatives
$\partial E^{\rm MF}/\partial \hat{Q}$
and
$\partial E^{\rm MF}/\partial \hat{Q}^\dag$.
For this goal, we expand
Eq.~(\ref{eq::E_MF_QQ})
in powers of
$\hat{Q}^\dag \hat{Q}$.
Using definition~(\ref{eq::matrix_diff_defin}),
we can demonstrate that
\begin{eqnarray}
\frac{\partial}{\partial \hat{Q}^\dag} {\rm Tr}\/ (\hat{Q}^\dag \hat{Q})^n
=
n \hat{Q} (\hat{Q}^\dag \hat{Q})^{n-1}.
\end{eqnarray}
We see that the differentiation rule for this monomial is essentially
identical to the rule for differentiating a product of commuting variables.
Such a simplification occurs due to invariance of the trace under cyclic
permutation of multipliers under sign of the trace. This allows us to perform
a re-summation of the power series and derive
\begin{eqnarray}
-\frac{\partial E^{\rm MF}}{\partial \hat{Q}^\dag}
=
\frac{1}{2N_c} \sum_{\bf q}
	\hat{Q}
	\left(
		\varepsilon_{\bf q}^2 + \hat{Q}^\dag \hat{Q}
	\right)^{-\frac{1}{2}}.
\end{eqnarray}
Thus we obtain the self-consistency
condition~(\ref{eq::hellm_feyn}).
Similarly, one can derive
\begin{eqnarray}
-\frac{\partial E^{\rm MF}}{\partial \hat{Q}}
=
\frac{1}{2N_c} \sum_{\bf q}
	\left(
		\varepsilon_{\bf q}^2 + \hat{Q}^\dag \hat{Q}
	\right)^{-\frac{1}{2}} \hat{Q}^\dag,
\end{eqnarray}
which can be used in
Eq.~(\ref{eq::Hell_Feyn_matrix2}).


\end{document}